\documentclass[reprint,amsmath,amssymb,prl,aps,superscriptaddress]{revtex4-1}
\pdfoutput=1

\usepackage[utf8x]{inputenc}

\usepackage{graphicx}
\usepackage{bm} % bold math
\usepackage{upgreek}
\usepackage{color}
\usepackage{url}
\usepackage{hyperref} % add hypertext capabilities
\hypersetup{
  final,
  pdfborder={0 0 0},
  pdftitle={Low temperature heat capacity of severely deformed
    metallic glass},
  pdfauthor={Bünz, Brink, Tsuchiya, Meng, Wilde, Albe},
  colorlinks=true,
  urlcolor=red,
  linkcolor=blue,
  citecolor=blue
}

\begin{document}

\title{Low temperature heat capacity of a severely deformed metallic glass}

\author{Jonas B\"unz}
\email{jonasbuenz@uni-muenster.de}
\affiliation{Institut f\"ur Materialphysik, Westf\"alische
  Wilhelms-Universit\"at M\"unster, Wilhelm-Klemm-Stra\ss{}e 10,
  D-48149 M\"unster, Germany}

\author{Tobias Brink}
\affiliation{Fachgebiet Materialmodellierung, Institut f{\"u}r
  Materialwissenschaft, TU Darmstadt,
  Jovanka-Bontschits-Stra\ss{}e~2, D-64287 Darmstadt, Germany}

\author{Koichi Tsuchiya}
\affiliation{National Institute of Materials Science, 1-2-1 Sengen,
  JP-305-0047 Tsukuba, Japan}

\author{Fanqiang Meng}
\affiliation{National Institute of Materials Science, 1-2-1 Sengen,
  JP-305-0047 Tsukuba, Japan}

\author{Gerhard Wilde}
\affiliation{Institut f\"ur Materialphysik, Westf\"alische
  Wilhelms-Universit\"at M\"unster, Wilhelm-Klemm-Stra\ss{}e 10,
  D-48149 M\"unster, Germany}

\author{Karsten Albe}
\affiliation{Fachgebiet Materialmodellierung, Institut f{\"u}r
  Materialwissenschaft, TU Darmstadt,
  Jovanka-Bontschits-Stra\ss{}e~2, D-64287 Darmstadt, Germany}

\date{April 1, 2014}

\begin{abstract}
  The low temperature heat capacity of amorphous materials reveals a
  low-frequency enhancement ({\em boson peak}) of the vibrational
  density of states, as compared with the Debye law.  By measuring the
  low-temperature heat capacity of a Zr-based bulk metallic glass
  relative to a crystalline reference state, we show that the heat
  capacity of the glass is strongly enhanced after severe plastic
  deformation by high-pressure torsion, while subsequent thermal
  annealing at elevated temperatures leads to a significant
  reduction. The detailed analysis of corresponding molecular dynamics
  simulations of an amorphous Zr-Cu glass shows that the change in
  heat capacity is primarily due to enhanced low-frequency modes
  within the shear band region.
  \vspace{0.25\baselineskip}\\
  {
    \noindent
    \footnotesize
    Published in:\\
    \href{http://journals.aps.org/prl/abstract/10.1103/PhysRevLett.112.135501}
         {J.~B\"unz \textit{et al.},
          Phys.\ Rev.\ Lett.\ \textbf{112}, 135501 (2014)}
    \hfill
    DOI: \href{http://dx.doi.org/10.1103/PhysRevLett.112.135501}
              {10.1103/PhysRevLett.112.135501}\\%
    \copyright{} 2014 American Physical Society.
  }
\end{abstract}

\maketitle

%\section{Introduction}

Metallic glasses show mechanical, electrical, and magnetical
properties, which can be very distinct from the properties of their
crystalline counterparts \cite{WangBulk2004}. One characteristic
feature of glassy materials is a low-frequency enhancement of the
vibrational density of states as compared with the Debye law.  This
excess contribution with respect to crystalline systems is commonly
referred to as the boson peak.  The boson peak is situated in the
terahertz region of the vibrational spectrum and hence also influences
the low temperature heat capacity.  After several decades of
controversy, it is now quite accepted that the boson peak is due to
quasi-localized transverse vibrational modes associated with
``defective'' soft local structures in the glass
\cite{Shintani2008,Schober2011501,Schober1991,Laird1991,Loeffler2012}.
The enhancement of the local vibrational density of states (often
referred to as ``soft modes'') leads to additional scattering of
phonons.
The resulting contribution to the heat capacity $c$ in excess of the
Debye $T^{3}$-law becomes visible as a peak, if $c(T)/T^{3}$ is
plotted in the temperature range of $T = 10\,\mathrm{K}$ to
$40\,\mathrm{K}$ \cite{Angell31031995}.  The boson peak is a general
feature of glasses, irrespective of the dominant type of the
interatomic or intermolecular binding energy and is thus also observed
in metallic glasses. Although there is agreement concerning the
connection between the phonon spectrum of glasses and the
peculiarities of parts of the spectrum related to the boson-peak
anomalies, the atomistic description of any structural origin of the
low-frequency excited states remains unclear despite of numerous
experimental and theoretical investigations. These excited states are
described in terms of soft harmonic potentials
\cite{karpov1983theory,gil1993low}, fluctuating force or elastic
constants
\cite{schirmacher1993vibrational,schirmacher1998harmonic,Schirmacher2006,Schirmacher2007,Ferrante2013},
strings of atoms \cite{schober1996low}, smeared out Van Hove
singularities \cite{Taraskin2001,Chumakov2011}, or interstitialcy-like
``defects'' \cite{vasiliev2009relationship}.  Even though the various
models for describing the boson peak behavior differ in their specific
assumptions and interpretations, they have in common that the glass
state is described by spatially heterogeneously distributed regions
having decreased elastic constants.  These regions are distributed
mesoscopically, i.e., on a scale of a few nanometers.  In metallic
glasses, rather localized and structurally disturbed regions occur
upon application of external stress exceeding the elastic limit. These
so-called shear bands have a typical thickness of the order of
$10\,\mathrm{nm}$ \cite{wilde2011nanocrystallization} and present the
regions where plastic shear strain was localized and quasi-plastic
straining occurred under the conditions of shear softening.

In order to analyze the contribution of shear bands to the
low-temperature heat capacity, we performed measurements on a
$\mathrm{Zr_{50}Cu_{40}Al_{10}}$ metallic glass that was subjected to
severe torsional deformation under a high applied hydrostatic pressure
\cite{Meng2012}.  In a previous investigation of that material,
individual shear bands could not be identified after applying large
strain to the sample \cite{Meng2012}.  Nonetheless, the sample
exhibits regions with deformation-induced structural changes, that can
be considered as ``macroscopic shear bands''.
Zr-based metallic glasses, especially the composition
investigated here, show rather ``strong'' behavior concerning their
fragility characteristics, which according to Ediger \textit{et al.},
should lead to a more pronounced boson-peak
\cite{ediger1996supercooled}. The sample was forced to withstand
catastrophic failure by crack formation despite the large strain via
geometrical confinement. In consequence, the regions of strong deformation
localization, i.e.\ the shear bands, experienced severe conditions and
the number of such regions is maximized \cite{Meng2012}.

Thus, this confined deformation up to severe strain values produces a
specific state that is well-suited to distinguish the contributions of shear band
and matrix regions outside the shear bands.  To that end we
performed molecular dynamics simulations of deformation of a
$\mathrm{Cu_{64}Zr_{36}}$ metallic glass---a model system comparable
to the Zr-Cu-Al glass---to differentiate the behavior of matrix and
shear band regions in the glass.

Furthermore, we measured the response of the heat capacity at low
temperatures
upon the thermomechanical processing of the glass to gain more
insights into the structural relaxation behavior of the deformed
metallic glass.

%\section{Experimental Details}

The sample material $\mathrm{Zr_{50}Cu_{40}Al_{10}}$ was cast from
ingots of pure metals to obtain an amorphous master ingot.  We cut the
master ingot into the specific dimensions for torsional deformation
under an applied hydrostatic pressure (HPT).  The HPT-discs had a
diameter of $10\,\mathrm{mm}$, and a thickness of
$0.89\,\mathrm{mm}$. We deformed the samples with an applied
pressure of $5\,\mathrm{GPa}$ for 50 revolutions. Further information
about the materials processing can be found in Ref. \onlinecite{Meng2012}.

The characteristic temperatures  were measured by differential
scanning calorimetry (DSC) experiments using a Perkin Elmer Diamond
DSC.  The experiments were performed with a constant heating rate of
$20\,\mathrm{K/min}$ under a high purity argon flow of
$20\,\mathrm{ml/min}$.

We take the glass transition temperature $T_\mathrm{g}$  as the
temperature where the increase of  heat capacity during the glass
transition amounts to half the value of the difference between the
heat capacity of the liquid phase $c_{p}^\mathrm{l}$ and the glass
phase $c_{p}^\mathrm{g}$. We measured a glass transition temperature of
$705\,\mathrm{K}$ and an onset of crystallization $T_\mathrm{x}$ at
$769\,\mathrm{K}$.  The undercooled liquid interval accessible to
calorimetry measurements at this heating rate amounts to $\Delta
T=T_\mathrm{x}-T_\mathrm{g}=64\,\mathrm{K}$.

The amorphicity of the samples was checked at room temperature by X-ray
diffraction (XRD) with a Siemens D5000 diffractometer. The samples,
both deformed and as-cast, show identical broad diffraction maxima
typical for the amorphous phase.  There is no sign of partial
crystallization, nor did the HPT deformation change the structure in a
way detectable in XRD.

The heat capacity $c_{p}$ was measured under high vacuum conditions
with a Quantum Design PPMS (Physical Property Measurement System),
which allows cooling down to $2\,\mathrm{K}$ with a liquid helium
system. The samples were polished for good thermal contact and cut
into $2.5 \times 2.5 \, \mathrm{mm}^2$ pieces, each with a mass of
$10\,\mathrm{mg}$ to $20\,\mathrm{mg}$. The samples were placed on top
of a sapphire block of known heat capacity $c_a$ with a thermal grease
to ensure good thermal contact. At each temperature, a known heat
pulse is applied and sample and platform are heated to a temperature
$T_{p}(t)$.  After the end of the heat pulse, the sample temperature
relaxes to the heat sink temperature $T_{0}$ according to
\begin{equation}
T_{p}(t)=T_{0}+\Delta T\exp(-t/\tau)
\end{equation}
with the time constant $\tau=(c_{p}+c_{a})/K$
\cite{Lashley2003369}. The factor $K$ is the thermal conductance between sample
platform and heat sink. The heat capacity $c_{p}$ can now be measured by fitting the
temperature decay $\tau$ with an accuracy better than 3\%. Prior to
the measurement of the samples, we measured $K$ and $c_{a}$ with an
empty sapphire crystal with the applied grease for baseline
correction.

Each heat capacity curve in this work includes three individual
measurements at each temperature to ensure reproducibility. We chose
the spacing between temperatures in a logarithmic way to
preferentially sample the region of low temperature.

The annealing of the samples above $394\,\mathrm{K}$ was done in a TA
Instruments Q500 DSC under constant argon flow of
$50\,\mathrm{ml/min}$.  Annealing at temperatures below
$393\,\mathrm{K}$ and for long times was performed in a Thermometric
TAM III microcalorimeter.  Both methods allow to simultaneously
measure the emitted heat flow to investigate relaxational events and
exclude crystallization.  For obtaining a crystalline reference of
identical composition, we heated a Zr$_{50}$Cu$_{40}$Al$_{10}$ sample
up to $823\,\mathrm{K}$, $60\,\mathrm{K}$ above the crystallization
temperature.

%\section{Simulation Methods}

Molecular dynamics (MD) computer simulations using the software
\textsc{lammps} \cite{Plimpton1995} were used for discriminating the
matrix and shear band contributions to the heat capacity.  For this
purpose, we simulated a $\mathrm{Cu_{64}Zr_{36}}$ system using the
Finnis-Sinclair type potential by Mendelev \textit{et al.}\
\cite{Mendelev2009}.  A bulk metallic glass sample was prepared by
quenching from the melt to $50\,\mathrm{K}$ with a cooling rate of
$10^{10}\,\mathrm{K/s}$.  The simulation box consisting of 364500
atoms was generated by a $2 \times 2 \times 2$ replication of the
quenched system.  We induced shear band formation using a volume
conserving pure shear deformation with a shear rate of
$10^8\,\mathrm{s^{-1}}$ up to a maximum shear of 20\%.  Additionally,
we annealed the deformed sample at $500\,\mathrm{K}$ for
$20\,\mathrm{ns}$ and $40\,\mathrm{ns}$.

We used the von Mises local shear invariant $\eta_i$
\cite{Shimizu2007} as implemented by \textsc{Ovito}
\cite{Stukowski2010} as a criterion for distinguishing atoms: Atoms
with $\eta_i > 0.2$ were marked as belonging to the shear band or
shear transformation zones (STZs), while all others were considered as
matrix atoms.

As a first step to obtain heat capacity values, we calculated the
phonon density of states (PDOS) from the Fourier-transform of the
velocity autocorrelation function \cite{Dickey1969}. This was done
separately for the atoms in the shear band and in the matrix after
unloading and equilibrating the system with a barostat at zero stress.

To calculate the corresponding heat capacities from the phonon density
of states $g(\omega)$, we employed the harmonic approximation of the
free energy \cite{Pathria1996}:
\begin{equation}
  c(T) =  k_\mathrm{B} \int_0^\infty
  \left( \frac{\hbar \omega}{2 k_\mathrm{B} T} \right)^2
  \sinh^{-2} \left( \frac{\hbar \omega}{2 k_\mathrm{B} T} \right)
  g(\omega) \mathrm{d}\omega.
\end{equation}
In this model we assume $c = c_V = c_p$, which gives a reasonable
approximation at low temperatures.  As a reference, we also calculated
heat capacities for the CuZr B2 crystal using the same method.

Note that the direct calculation of the specific heat from the time
averaged inner energy given by the classical trajectories in MD simulations
would yield unphysical values in the low temperature regime due to the missing
quantum-statistical contributions.

%\section{Results}

\begin{figure}
  \centering
  \includegraphics[width=8.6cm]{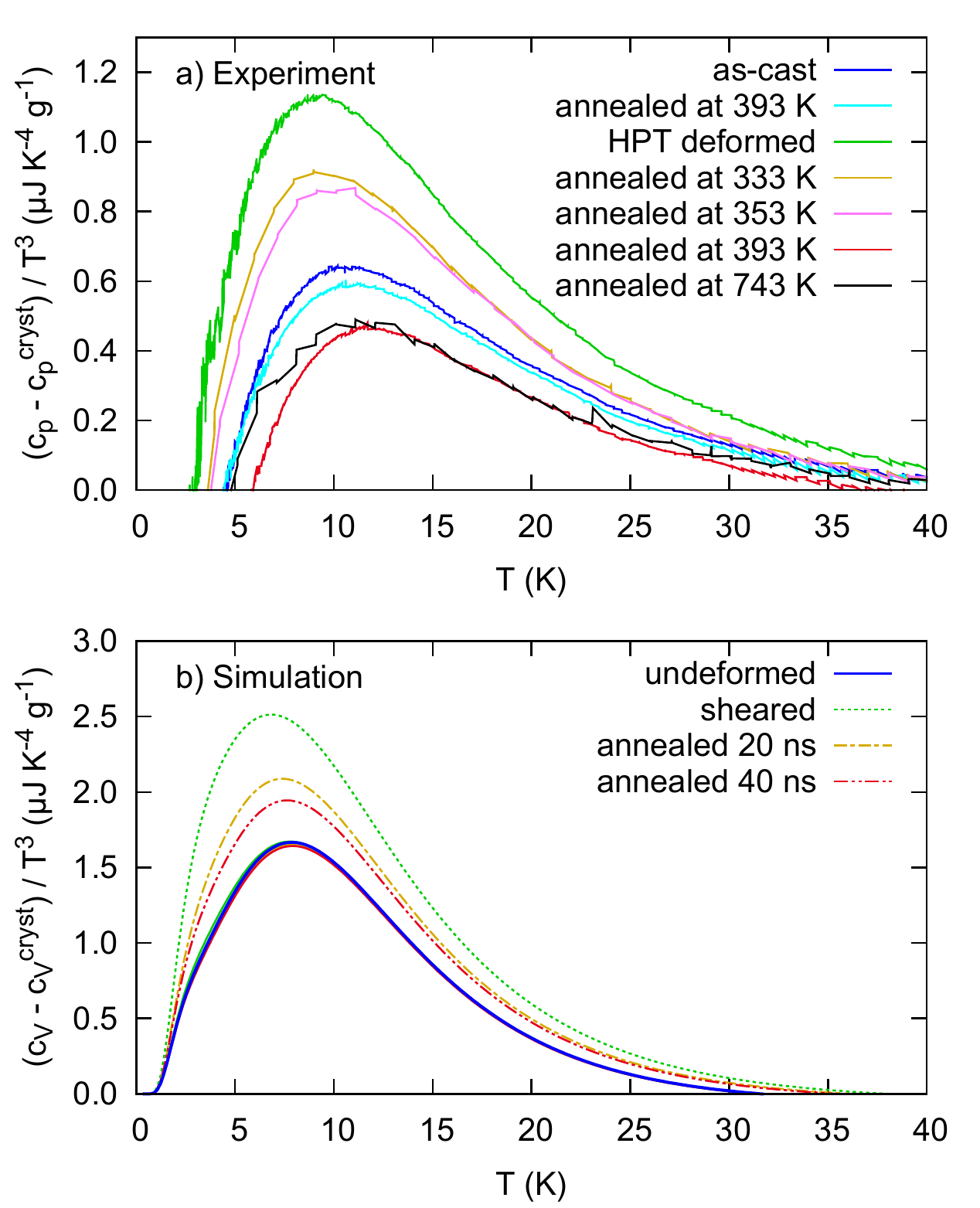}
  \caption{Difference in heat capacity between glass and crystal. a)
    Experimental data, annealing below $393\,\mathrm{K}$ was done for
    one week, annealing at $743\,\mathrm{K}$ for $10\,\mathrm{s}$. b)
    Simulation data, dashed lines are data from shear bands, solid
    lines from the rest of the sample.}
  \label{fig:cV}
\end{figure}

The experimentally measured heat capacity is plotted as $\Delta
c_{p}/T^{3}$ versus $T$ in Fig.~\ref{fig:cV}(a), making visible the
contribution of the boson peak at low temperatures. The $c_{p}$-curve
of the crystalline Zr-Cu-Al sample is used as reference and therefore
subtracted from the data of each amorphous sample.  Since the
crystalline sample shows no contribution to the boson peak but a
comparable heat capacity from electronic and ``lattice''
contributions, the excess heat capacity plotted in
Fig.~\ref{fig:cV}(a) is solely due the boson peak.

The HPT-deformed glass shows an enhanced peak, whose height is almost
twice as high compared to the as-cast state.  The peak maximum of the
deformed sample is slightly shifted by $2.7\,\mathrm{K}$ to lower
temperatures.

We annealed the samples at different temperatures with simultaneous
measurement of the relaxational heat flow in the
microcalorimeter. After at most seven days the heat flow signal
reached the background, indicating that the sample was close to an
equilibrium state at that temperature. The undeformed sample shows a
slight decrease of the boson peak after annealing at
$393\,\mathrm{K}$, while annealing at $333\,\mathrm{K}$ and
$353\,\mathrm{K}$ lowers the boson peak of the deformed sample to an
intermediate height between the deformed and the as-cast state. The
deformation ``defects'' are not completely healed by the thermal
treatment.  Otherwise the sample that was annealed at
$393\,\mathrm{K}$ shows a boson peak which is even lower than the
boson peak of the as-cast sample or of the undeformed sample relaxed
at $393\,\mathrm{K}$. Additional DSC experiments were performed and
did not show any evidence for partial crystallization during
annealing.  Furthermore, we heated a HPT-deformed sample to
$743\,\mathrm{K}>T_\mathrm{g}$. The sample reaches internal
equilibrium within the first $10\,\mathrm{s}$ of annealing and has a
defined thermal history after cooling with $20\,\mathrm{K/min}$. The
boson peak has the same height as the one of the HPT-deformed samples
which is in metastable equilibrium at $393\,\mathrm{K}$. The
undeformed sample, which was equilibrated at the same temperature,
shows a $20\%$ higher boson peak.

\begin{figure}
  \centering
  \includegraphics[width=8.6cm]{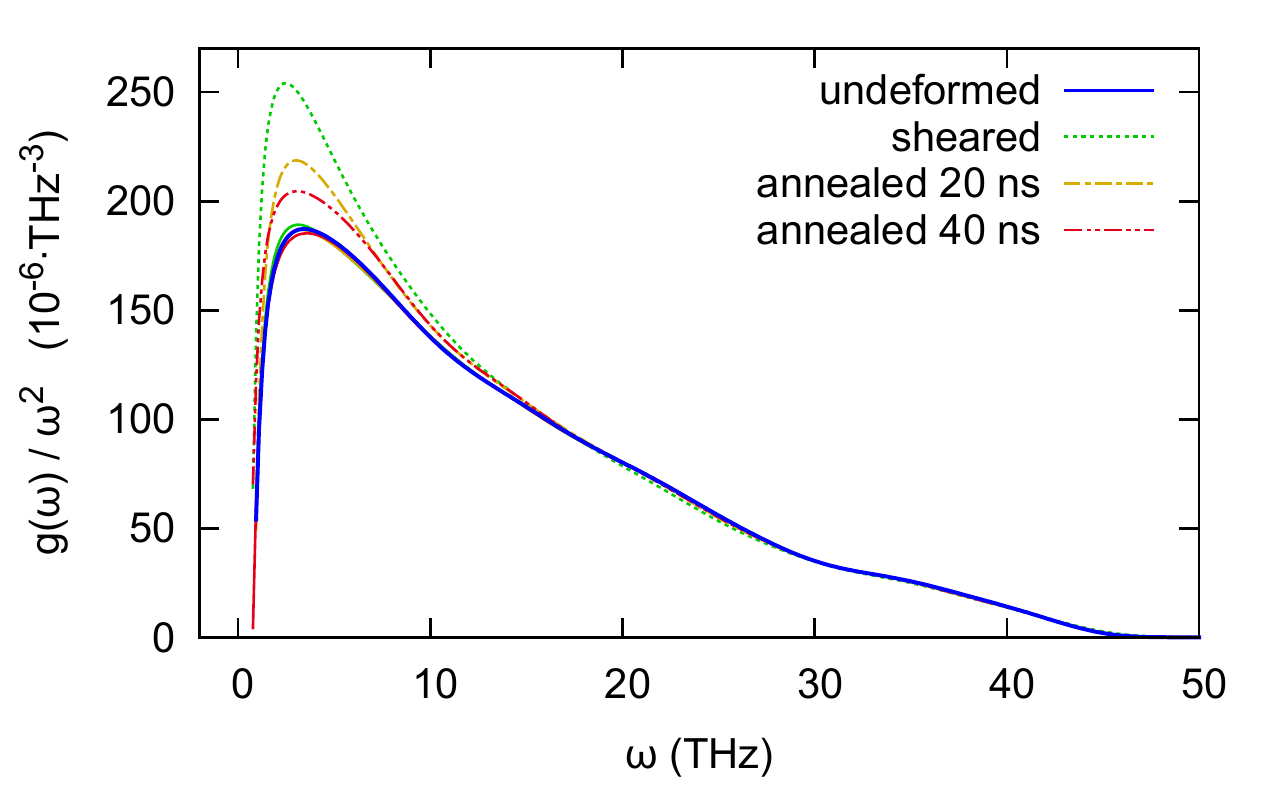}
\caption{Phonon density of states from computer simulation.  Dashed
  lines are data from shear bands, solid lines from the rest of the
  sample.}
  \label{fig:PDOS}
\end{figure}

In Figure~\ref{fig:PDOS} the PDOS calculated by the computer
simulation is shown.  The data is split into contributions of the
matrix (solid lines) and contributions of the shear band (dashed
lines).  The data is plotted for the as-quenched glass, the glass
deformed to a shear of 20\%, and the deformed sample annealed for
$20\,\mathrm{ns}$ and $40\,\mathrm{ns}$ at $500\,\mathrm{K}$.  Most
evidently, the vibrational spectrum of the matrix does not change
between the undeformed, sheared, and annealed samples, whereas the
spectrum of the shear band shows an increased boson peak which reduces
again with annealing.  Figure \ref{fig:cV}(b) shows the corresponding
heat capacities $c_V$ calculated from the PDOS.  As in the
experiment---and as already suggested by the PDOS---the boson peak in
the heat capacity increases after deformation and decreases again
after annealing. The simulation overestimates the difference in heat
capacity between crystal and glass but shows qualitatively the same
features as the experiment.  Due to the small timescale accessible by
the simulation we don't observe a complete recovery of the shear band
in the deformed glass by thermal annealing.

However, the computer simulations clearly show that the shear band
regions relax to a state of lower excess heat capacity, while the
contribution of the matrix is not changing.  The experimentally
observed fact that the deformed sample shows a significantly reduced
boson peak after thermal annealing at high temperature as compared to
the as-cast case is indicative for a structural relaxation.  Since the
excess with respect to the crystalline reference is reduced, this
suggests a relaxation towards a more crystal-like state within the
shear band region.  This interpretation is supported by recent results
indicating the prevalence of shear bands for the formation of
nanocrystals or crystal-like medium range order
\cite{wilde2011nanocrystallization,RoesnerShearBands}.  The enhanced
atomic mobility within the shear bands \cite{PhysRevLett.107.235503}
does also favor the relaxation of the modified glass structure towards
a more ordered state.

The change of the boson peak with applied pressure can often be
explained by mere changes of the elastic constants.  In these cases
the change of the peak height is accompanied by a shift of the
boson-peak frequency. When plotting the data in units of the Debye
frequency, the peaks are identical
\cite{Schirmacher2006,Chumakov2011}.  In our case the frequency of the
boson peak, and therefore its position in the heat capacity plot, does
not shift in either experiment or simulation.  This indicates a real
change of the disorder statistics, i.e., the distribution of local
moduli, and not only a change of the global elasticity.

All in all, our findings clearly show that all changes of the specific
heat induced by plastic deformation and annealing stem from the
regions of strain localization.  The matrix will stay relatively
intact in the processes.

%\section{Conclusion}

Both experimental and simulation data show that plastic deformation
enhances the boson contribution to the heat capacity.  Molecular
dynamics simulations provide direct
evidence that the additional contributions to the boson peak result
from structures that are located in the shear bands, while the PDOS in
the matrix stays unchanged with deformation and annealing. The
experiment supports this perspective as the change of boson peak
height of the undeformed samples during annealing is relatively small
compared to the deformed samples.

It should be noted that in our current work the experimental
conditions always lead to a strong strain localization. The analysis
given here does not cover a more homogeneous deformation behavior, in
which the matrix could also show a changed vibrational spectrum, if
there is even a distinction between shear band and matrix.  In this
case an increase of the boson peak after deformation may be observed
over the whole system.

Several studies show that shear bands can heal during annealing even
at temperatures well below $T_{g}$
\cite{Jiang20053469,Xie20085202,Ritter20117082}.  In contrast, our
findings indicate that the shear band structure after annealing
differs from an annealed, undeformed glass.  Diffusion studies
demonstrate accelerated diffusion \cite{PhysRevLett.107.235503} and
excess free volume \cite{Kanungo20041073} inside of shear bands.  This
could promote structural relaxation and short to medium range ordering
that drives the system to another meta-basin in the energy landscape
\cite{0953-8984-20-37-373101}. This meta-basin would exhibit different
structural properties: The decrease of the boson peak below the values
of the undeformed glass in that case suggests an increasingly
crystal-like structure.

%\section{Acknowledgements}

The authors thank Yoshihiko Yokayama (Institute of Metal Research,
Tohoku University) for providing the amorphous ingot material and
Leonie Koch for providing the model structures.
The authors gratefully acknowledge the computing time granted by the
John von Neumann Institute for Computing (NIC) and provided on the
supercomputer JUROPA at J\"ulich Supercomputing Centre (JSC).
Financial support by DFG grants no.\ Al-578/13-1 and WI-1899/12-1 is
gratefully acknowledged. This work is partly supported by the
Grant-in-Aid for Scientific Research on Innovative Area ``Bulk
Nanostructured Metals'' through MEXT, Japan (contract no.\ 22102004).

%\bibliographystyle{apsrev4-1}
%\bibliography{literatur}

\begin{thebibliography}{35}%
\makeatletter
\providecommand \@ifxundefined [1]{%
 \@ifx{#1\undefined}
}%
\providecommand \@ifnum [1]{%
 \ifnum #1\expandafter \@firstoftwo
 \else \expandafter \@secondoftwo
 \fi
}%
\providecommand \@ifx [1]{%
 \ifx #1\expandafter \@firstoftwo
 \else \expandafter \@secondoftwo
 \fi
}%
\providecommand \natexlab [1]{#1}%
\providecommand \enquote  [1]{``#1''}%
\providecommand \bibnamefont  [1]{#1}%
\providecommand \bibfnamefont [1]{#1}%
\providecommand \citenamefont [1]{#1}%
\providecommand \href@noop [0]{\@secondoftwo}%
\providecommand \href [0]{\begingroup \@sanitize@url \@href}%
\providecommand \@href[1]{\@@startlink{#1}\@@href}%
\providecommand \@@href[1]{\endgroup#1\@@endlink}%
\providecommand \@sanitize@url [0]{\catcode `\\12\catcode `\$12\catcode
  `\&12\catcode `\#12\catcode `\^12\catcode `\_12\catcode `\%12\relax}%
\providecommand \@@startlink[1]{}%
\providecommand \@@endlink[0]{}%
\providecommand \url  [0]{\begingroup\@sanitize@url \@url }%
\providecommand \@url [1]{\endgroup\@href {#1}{\urlprefix }}%
\providecommand \urlprefix  [0]{URL }%
\providecommand \Eprint [0]{\href }%
\providecommand \doibase [0]{http://dx.doi.org/}%
\providecommand \selectlanguage [0]{\@gobble}%
\providecommand \bibinfo  [0]{\@secondoftwo}%
\providecommand \bibfield  [0]{\@secondoftwo}%
\providecommand \translation [1]{[#1]}%
\providecommand \BibitemOpen [0]{}%
\providecommand \bibitemStop [0]{}%
\providecommand \bibitemNoStop [0]{.\EOS\space}%
\providecommand \EOS [0]{\spacefactor3000\relax}%
\providecommand \BibitemShut  [1]{\csname bibitem#1\endcsname}%
\let\auto@bib@innerbib\@empty
%</preamble>
\bibitem [{\citenamefont {Wang}\ \emph {et~al.}(2004)\citenamefont {Wang},
  \citenamefont {Dong},\ and\ \citenamefont {Shek}}]{WangBulk2004}%
  \BibitemOpen
  \bibfield  {author} {\bibinfo {author} {\bibfnamefont {W.~H.}\ \bibnamefont
  {Wang}}, \bibinfo {author} {\bibfnamefont {C.}~\bibnamefont {Dong}},\ and\
  \bibinfo {author} {\bibfnamefont {C.~H.}\ \bibnamefont {Shek}},\ }\href
  {http://dx.doi.org/10.1016/j.mser.2004.03.001} {\bibfield  {journal}
  {\bibinfo  {journal} {Mat. Sci. Eng. R}\ }\textbf {\bibinfo {volume} {44}}
  (\bibinfo {year} {2004})}\BibitemShut {NoStop}%
\bibitem [{\citenamefont {Shintani}\ and\ \citenamefont
  {Tanaka}(2008)}]{Shintani2008}%
  \BibitemOpen
  \bibfield  {author} {\bibinfo {author} {\bibfnamefont {H.}~\bibnamefont
  {Shintani}}\ and\ \bibinfo {author} {\bibfnamefont {H.}~\bibnamefont
  {Tanaka}},\ }\href {http://dx.doi.org/10.1038/nmat2293} {\bibfield  {journal}
  {\bibinfo  {journal} {Nat. Mater.}\ }\textbf {\bibinfo {volume} {7}},\
  \bibinfo {pages} {870} (\bibinfo {year} {2008})}\BibitemShut {NoStop}%
\bibitem [{\citenamefont {Schober}(2011)}]{Schober2011501}%
  \BibitemOpen
  \bibfield  {author} {\bibinfo {author} {\bibfnamefont {H.~R.}\ \bibnamefont
  {Schober}},\ }\href {http://dx.doi.org/10.1016/j.jnoncrysol.2010.07.036}
  {\bibfield  {journal} {\bibinfo  {journal} {J. Non-Cryst. Solids}\ }\textbf
  {\bibinfo {volume} {357}},\ \bibinfo {pages} {501} (\bibinfo {year}
  {2011})}\BibitemShut {NoStop}%
\bibitem [{\citenamefont {Schober}\ and\ \citenamefont
  {Laird}(1991)}]{Schober1991}%
  \BibitemOpen
  \bibfield  {author} {\bibinfo {author} {\bibfnamefont {H.~R.}\ \bibnamefont
  {Schober}}\ and\ \bibinfo {author} {\bibfnamefont {B.~B.}\ \bibnamefont
  {Laird}},\ }\href {\doibase 10.1103/PhysRevB.44.6746} {\bibfield  {journal}
  {\bibinfo  {journal} {Phys. Rev. B}\ }\textbf {\bibinfo {volume} {44}},\
  \bibinfo {pages} {6746} (\bibinfo {year} {1991})}\BibitemShut {NoStop}%
\bibitem [{\citenamefont {Laird}\ and\ \citenamefont
  {Schober}(1991)}]{Laird1991}%
  \BibitemOpen
  \bibfield  {author} {\bibinfo {author} {\bibfnamefont {B.~B.}\ \bibnamefont
  {Laird}}\ and\ \bibinfo {author} {\bibfnamefont {H.~R.}\ \bibnamefont
  {Schober}},\ }\href {\doibase 10.1103/PhysRevLett.66.636} {\bibfield
  {journal} {\bibinfo  {journal} {Phys. Rev. Lett.}\ }\textbf {\bibinfo
  {volume} {66}},\ \bibinfo {pages} {636} (\bibinfo {year} {1991})}\BibitemShut
  {NoStop}%
\bibitem [{\citenamefont {Derlet}\ \emph {et~al.}(2012)\citenamefont {Derlet},
  \citenamefont {Maa{\ss}},\ and\ \citenamefont {L\"offler}}]{Loeffler2012}%
  \BibitemOpen
  \bibfield  {author} {\bibinfo {author} {\bibfnamefont {P.~M.}\ \bibnamefont
  {Derlet}}, \bibinfo {author} {\bibfnamefont {R.}~\bibnamefont {Maa{\ss}}},\
  and\ \bibinfo {author} {\bibfnamefont {J.~F.}\ \bibnamefont {L\"offler}},\
  }\href {\doibase 10.1140/epjb/e2012-20902-0} {\bibfield  {journal} {\bibinfo
  {journal} {Eur. Phys. J. B}\ }\textbf {\bibinfo {volume} {85}},\ \bibinfo
  {pages} {1} (\bibinfo {year} {2012})}\BibitemShut {NoStop}%
\bibitem [{\citenamefont {Angell}(1995)}]{Angell31031995}%
  \BibitemOpen
  \bibfield  {author} {\bibinfo {author} {\bibfnamefont {C.~A.}\ \bibnamefont
  {Angell}},\ }\href {http://dx.doi.org/10.1126/science.267.5206.1924}
  {\bibfield  {journal} {\bibinfo  {journal} {Science}\ }\textbf {\bibinfo
  {volume} {267}},\ \bibinfo {pages} {1924} (\bibinfo {year}
  {1995})}\BibitemShut {NoStop}%
\bibitem [{\citenamefont {Karpov}\ \emph {et~al.}(1983)\citenamefont {Karpov},
  \citenamefont {Klinger},\ and\ \citenamefont {Ignat'ev}}]{karpov1983theory}%
  \BibitemOpen
  \bibfield  {author} {\bibinfo {author} {\bibfnamefont {V.~G.}\ \bibnamefont
  {Karpov}}, \bibinfo {author} {\bibfnamefont {M.~I.}\ \bibnamefont {Klinger}},\
  and\ \bibinfo {author} {\bibfnamefont {F.~N.}\ \bibnamefont {Ignat'ev}},\
  }\href {http://jetp.ac.ru/cgi-bin/e/index/e/57/2/p439?a=list} {\bibfield
  {journal} {\bibinfo  {journal} {Zh. Eksp. Teor. Fiz.}\ }\textbf {\bibinfo
  {volume} {84}},\ \bibinfo {pages} {760} (\bibinfo {year} {1983})}\BibitemShut
  {NoStop}%
\bibitem [{\citenamefont {Gil}\ \emph {et~al.}(1993)\citenamefont {Gil},
  \citenamefont {Ramos}, \citenamefont {Bringer},\ and\ \citenamefont
  {Buchenau}}]{gil1993low}%
  \BibitemOpen
  \bibfield  {author} {\bibinfo {author} {\bibfnamefont {L.}~\bibnamefont
  {Gil}}, \bibinfo {author} {\bibfnamefont {M.~A.}\ \bibnamefont {Ramos}},
  \bibinfo {author} {\bibfnamefont {A.}~\bibnamefont {Bringer}},\ and\
  \bibinfo{author}{\bibfnamefont{U.}~\bibnamefont{Buchenau}},\ }\href
  {http://dx.doi.org/10.1103/PhysRevLett.70.182} {\bibfield  {journal}
  {\bibinfo  {journal} {Phys. Rev. Lett.}\ }\textbf {\bibinfo {volume} {70}},\
  \bibinfo {pages} {182} (\bibinfo {year} {1993})}\BibitemShut {NoStop}%
\bibitem [{\citenamefont {Schirmacher}\ and\ \citenamefont
  {Wagener}(1993)}]{schirmacher1993vibrational}%
  \BibitemOpen
  \bibfield  {author} {\bibinfo {author} {\bibfnamefont {W.}~\bibnamefont
  {Schirmacher}}\ and\ \bibinfo {author} {\bibfnamefont {M.}~\bibnamefont
  {Wagener}},\ }\href {http://dx.doi.org/10.1016/0038-1098(93)90147-F}
  {\bibfield  {journal} {\bibinfo  {journal} {Solid State Commun.}\ }\textbf
  {\bibinfo {volume} {86}},\ \bibinfo {pages} {597} (\bibinfo {year}
  {1993})}\BibitemShut {NoStop}%
\bibitem [{\citenamefont {Schirmacher}\ \emph {et~al.}(1998)\citenamefont
  {Schirmacher}, \citenamefont {Diezemann},\ and\ \citenamefont
  {Ganter}}]{schirmacher1998harmonic}%
  \BibitemOpen
  \bibfield  {author} {\bibinfo {author} {\bibfnamefont {W.}~\bibnamefont
  {Schirmacher}}, \bibinfo {author} {\bibfnamefont {G.}~\bibnamefont
  {Diezemann}},\ and\ \bibinfo {author} {\bibfnamefont {C.}~\bibnamefont
  {Ganter}},\ }\href {http://dx.doi.org/10.1103/PhysRevLett.81.136} {\bibfield
  {journal} {\bibinfo  {journal} {Phys. Rev. Lett.}\ }\textbf {\bibinfo
  {volume} {81}},\ \bibinfo {pages} {136} (\bibinfo {year} {1998})}\BibitemShut
  {NoStop}%
\bibitem [{\citenamefont {Schirmacher}(2006)}]{Schirmacher2006}%
  \BibitemOpen
  \bibfield  {author} {\bibinfo {author} {\bibfnamefont {W.}~\bibnamefont
  {Schirmacher}},\ }\href {\doibase 10.1209/epl/i2005-10471-9} {\bibfield
  {journal} {\bibinfo  {journal} {Europhys. Lett.}\ }\textbf {\bibinfo {volume}
  {73}},\ \bibinfo {pages} {892} (\bibinfo {year} {2006})}\BibitemShut
  {NoStop}%
\bibitem [{\citenamefont {Schirmacher}\ \emph {et~al.}(2007)\citenamefont
  {Schirmacher}, \citenamefont {Ruocco},\ and\ \citenamefont
  {Scopigno}}]{Schirmacher2007}%
  \BibitemOpen
  \bibfield  {author} {\bibinfo {author} {\bibfnamefont {W.}~\bibnamefont
  {Schirmacher}}, \bibinfo {author} {\bibfnamefont {G.}~\bibnamefont {Ruocco}},\
  and\ \bibinfo {author} {\bibfnamefont {T.}~\bibnamefont {Scopigno}},\
  }\href {\doibase 10.1103/PhysRevLett.98.025501} {\bibfield  {journal}
  {\bibinfo  {journal} {Phys. Rev. Lett.}\ }\textbf {\bibinfo {volume} {98}},\
  \bibinfo {pages} {025501} (\bibinfo {year} {2007})}\BibitemShut {NoStop}%
\bibitem [{\citenamefont {Ferrante}\ \emph {et~al.}(2013)\citenamefont
  {Ferrante}, \citenamefont {Pontecorvo}, \citenamefont {Cerullo},
  \citenamefont {Chiasera}, \citenamefont {Ruocco}, \citenamefont
  {Schirmacher},\ and\ \citenamefont {Scopigno}}]{Ferrante2013}%
  \BibitemOpen
  \bibfield  {author} {\bibinfo {author} {\bibfnamefont {C.}~\bibnamefont
  {Ferrante}}, \bibinfo {author} {\bibfnamefont {W.}~\bibnamefont
  {Pontecorvo}}, \bibinfo {author} {\bibfnamefont {G.}~\bibnamefont {Cerullo}},
  \bibinfo {author} {\bibfnamefont {A.}~\bibnamefont {Chiasera}}, \bibinfo
  {author} {\bibfnamefont {G.}~\bibnamefont {Ruocco}}, \bibinfo {author}
  {\bibfnamefont {W.}~\bibnamefont {Schirmacher}},\ and\ \bibinfo {author}
  {\bibfnamefont {T.}~\bibnamefont {Scopigno}},\ }\href {\doibase
  10.1038/ncomms2826} {\bibfield  {journal} {\bibinfo  {journal} {Nat.
  Commun.}\ }\textbf {\bibinfo {volume} {4}},\ \bibinfo {pages} {1793}
  (\bibinfo {year} {2013})}\BibitemShut {NoStop}%
\bibitem [{\citenamefont {Schober}\ and\ \citenamefont
  {Oligschleger}(1996)}]{schober1996low}%
  \BibitemOpen
  \bibfield  {author} {\bibinfo {author} {\bibfnamefont {H.~R.}\ \bibnamefont
  {Schober}}\ and\ \bibinfo {author} {\bibfnamefont {C.}~\bibnamefont
  {Oligschleger}},\ }\href {http://dx.doi.org/10.1103/PhysRevB.53.11469}
  {\bibfield  {journal} {\bibinfo  {journal} {Phys. Rev. B}\ }\textbf {\bibinfo
  {volume} {53}},\ \bibinfo {pages} {11469} (\bibinfo {year}
  {1996})}\BibitemShut {NoStop}%
\bibitem [{\citenamefont {Taraskin}\ \emph {et~al.}(2001)\citenamefont
  {Taraskin}, \citenamefont {Loh}, \citenamefont {Natarajan},\ and\
  \citenamefont {Elliott}}]{Taraskin2001}%
  \BibitemOpen
  \bibfield  {author} {\bibinfo {author} {\bibfnamefont {S.~N.}\ \bibnamefont
  {Taraskin}}, \bibinfo {author} {\bibfnamefont {Y.~L.}\ \bibnamefont {Loh}},
  \bibinfo {author} {\bibfnamefont {G.}~\bibnamefont {Natarajan}},\ and\
  \bibinfo {author} {\bibfnamefont {S.~R.}\ \bibnamefont {Elliott}},\ }\href
  {http://dx.doi.org/10.1103/PhysRevLett.86.1255} {\bibfield  {journal}
  {\bibinfo  {journal} {Phys. Rev. Lett.}\ }\textbf {\bibinfo {volume} {86}},\
  \bibinfo {pages} {1255} (\bibinfo {year} {2001})}\BibitemShut {NoStop}%
\bibitem [{\citenamefont {Chumakov}\ \emph {et~al.}(2011)\citenamefont
  {Chumakov}, \citenamefont {Monaco}, \citenamefont {Monaco}, \citenamefont
  {Crichton}, \citenamefont {Bosak}, \citenamefont {R\"uffer}, \citenamefont
  {Meyer}, \citenamefont {Kargl}, \citenamefont {Comez}, \citenamefont
  {Fioretto}, \citenamefont {Giefers}, \citenamefont {Roitsch}, \citenamefont
  {Wortmann}, \citenamefont {Manghnani}, \citenamefont {Hushur}, \citenamefont
  {Williams}, \citenamefont {Balogh}, \citenamefont
  {Parli\ifmmode~\acute{n}\else \'{n}\fi{}ski}, \citenamefont {Jochym},\ and\
  \citenamefont {Piekarz}}]{Chumakov2011}%
  \BibitemOpen
  \bibfield  {author} {\bibinfo {author} {\bibfnamefont {A.~I.}\ \bibnamefont
  {Chumakov}}, \bibinfo {author} {\bibfnamefont {G.}~\bibnamefont {Monaco}},
  \bibinfo {author} {\bibfnamefont {A.}~\bibnamefont {Monaco}}, \bibinfo
  {author} {\bibfnamefont {W.~A.}\ \bibnamefont {Crichton}}, \bibinfo {author}
  {\bibfnamefont {A.}~\bibnamefont {Bosak}}, \bibinfo {author} {\bibfnamefont
  {R.}~\bibnamefont {R\"uffer}}, \bibinfo {author} {\bibfnamefont
  {A.}~\bibnamefont {Meyer}}, \bibinfo {author} {\bibfnamefont
  {F.}~\bibnamefont {Kargl}}, \bibinfo {author} {\bibfnamefont
  {L.}~\bibnamefont {Comez}}, \bibinfo {author} {\bibfnamefont
  {D.}~\bibnamefont {Fioretto}}, \bibinfo {author} {\bibfnamefont
  {H.}~\bibnamefont {Giefers}}, \bibinfo {author} {\bibfnamefont
  {S.}~\bibnamefont {Roitsch}}, \bibinfo {author} {\bibfnamefont
  {G.}~\bibnamefont {Wortmann}}, \bibinfo {author} {\bibfnamefont {M.~H.}\
  \bibnamefont {Manghnani}}, \bibinfo {author} {\bibfnamefont {A.}~\bibnamefont
  {Hushur}}, \bibinfo {author} {\bibfnamefont {Q.}~\bibnamefont {Williams}},
  \bibinfo {author} {\bibfnamefont {J.}~\bibnamefont {Balogh}}, \bibinfo
  {author} {\bibfnamefont {K.}~\bibnamefont {Parli\ifmmode~\acute{n}\else
  \'{n}\fi{}ski}}, \bibinfo {author} {\bibfnamefont {P.}~\bibnamefont
  {Jochym}},\ and\ \bibinfo {author} {\bibfnamefont {P.}~\bibnamefont
  {Piekarz}},\ }\href {http://dx.doi.org/10.1103/PhysRevLett.106.225501}
  {\bibfield  {journal} {\bibinfo  {journal} {Phys. Rev. Lett.}\ }\textbf
  {\bibinfo {volume} {106}},\ \bibinfo {pages} {225501} (\bibinfo {year}
  {2011})}\BibitemShut {NoStop}%
\bibitem [{\citenamefont {Vasiliev}\ \emph {et~al.}(2009)\citenamefont
  {Vasiliev}, \citenamefont {Voloshok}, \citenamefont {Granato}, \citenamefont
  {Joncich}, \citenamefont {Mitrofanov},\ and\ \citenamefont
  {Khonik}}]{vasiliev2009relationship}%
  \BibitemOpen
  \bibfield  {author} {\bibinfo {author} {\bibfnamefont {A.~N.}\ \bibnamefont
  {Vasiliev}}, \bibinfo {author} {\bibfnamefont {T.~N.}\ \bibnamefont
  {Voloshok}}, \bibinfo {author} {\bibfnamefont {A.~V.}\ \bibnamefont
  {Granato}}, \bibinfo {author} {\bibfnamefont {D.~M.}\ \bibnamefont
  {Joncich}}, \bibinfo {author} {\bibfnamefont {Y.~P.}\ \bibnamefont
  {Mitrofanov}},\ and\ \bibinfo {author} {\bibfnamefont {V.~A.}\ \bibnamefont
  {Khonik}},\ }\href {http://dx.doi.org/10.1103/PhysRevB.80.172102} {\bibfield
  {journal} {\bibinfo  {journal} {Phys. Rev. B}\ }\textbf {\bibinfo {volume}
  {80}},\ \bibinfo {pages} {172102} (\bibinfo {year} {2009})}\BibitemShut
  {NoStop}%
\bibitem [{\citenamefont {Wilde}\ and\ \citenamefont
  {R\"osner}(2011)}]{wilde2011nanocrystallization}%
  \BibitemOpen
  \bibfield  {author} {\bibinfo {author} {\bibfnamefont {G.}~\bibnamefont
  {Wilde}}\ and\ \bibinfo {author} {\bibfnamefont {H.}~\bibnamefont
  {R\"osner}},\ }\href {http://dx.doi.org/10.1063/1.3602315} {\bibfield
  {journal} {\bibinfo  {journal} {Appl. Phys. Lett.}\ }\textbf {\bibinfo
  {volume} {98}},\ \bibinfo {pages} {251904} (\bibinfo {year}
  {2011})}\BibitemShut {NoStop}%
\bibitem [{\citenamefont {Meng}\ \emph {et~al.}(2012)\citenamefont {Meng},
  \citenamefont {Tsuchiya}, \citenamefont {{Seiichiro II}},\ and\ \citenamefont
  {Yokoyama}}]{Meng2012}%
  \BibitemOpen
  \bibfield  {author} {\bibinfo {author} {\bibfnamefont {F.}~\bibnamefont
  {Meng}}, \bibinfo {author} {\bibfnamefont {K.}~\bibnamefont {Tsuchiya}},
  \bibinfo {author} {\bibnamefont {{Seiichiro II}}},\ and\ \bibinfo {author}
  {\bibfnamefont {Y.}~\bibnamefont {Yokoyama}},\ }\href
  {http://dx.doi.org/10.1063/1.4753998} {\bibfield  {journal} {\bibinfo
  {journal} {Appl. Phys. Lett.}\ }\textbf {\bibinfo {volume} {101}},\ \bibinfo
  {pages} {121914} (\bibinfo {year} {2012})}\BibitemShut {NoStop}%
\bibitem [{\citenamefont {Ediger}\ \emph {et~al.}(1996)\citenamefont {Ediger},
  \citenamefont {Angell},\ and\ \citenamefont {Nagel}}]{ediger1996supercooled}%
  \BibitemOpen
  \bibfield  {author} {\bibinfo {author} {\bibfnamefont {M.~D.}\ \bibnamefont
  {Ediger}}, \bibinfo {author} {\bibfnamefont {C.~A.}\ \bibnamefont {Angell}},\
  and\ \bibinfo {author} {\bibfnamefont {S.~R.}\ \bibnamefont {Nagel}},\
  }\href {http://dx.doi.org/10.1021/jp953538d} {\bibfield  {journal} {\bibinfo
  {journal} {J. Phys. Chem.}\ }\textbf {\bibinfo {volume} {100}},\ \bibinfo
  {pages} {13200} (\bibinfo {year} {1996})}\BibitemShut {NoStop}%
\bibitem [{\citenamefont {Lashley}\ \emph {et~al.}(2003)\citenamefont
  {Lashley}, \citenamefont {Hundley}, \citenamefont {Migliori}, \citenamefont
  {Sarrao}, \citenamefont {Pagliuso}, \citenamefont {Darling}, \citenamefont
  {Jaime}, \citenamefont {Cooley}, \citenamefont {Hults}, \citenamefont
  {Morales}, \citenamefont {Thoma}, \citenamefont {Smith}, \citenamefont
  {Boerio-Goates}, \citenamefont {Woodfield}, \citenamefont {Stewart},
  \citenamefont {Fisher},\ and\ \citenamefont {Phillips}}]{Lashley2003369}%
  \BibitemOpen
  \bibfield  {author} {\bibinfo {author} {\bibfnamefont {J.~C.}\ \bibnamefont
  {Lashley}}, \bibinfo {author} {\bibfnamefont {M.~F.}\ \bibnamefont
  {Hundley}}, \bibinfo {author} {\bibfnamefont {A.}~\bibnamefont {Migliori}},
  \bibinfo {author} {\bibfnamefont {J.~L.}\ \bibnamefont {Sarrao}}, \bibinfo
  {author} {\bibfnamefont {P.~G.}\ \bibnamefont {Pagliuso}}, \bibinfo {author}
  {\bibfnamefont {T.~W.}\ \bibnamefont {Darling}}, \bibinfo {author}
  {\bibfnamefont {M.}~\bibnamefont {Jaime}}, \bibinfo {author} {\bibfnamefont
  {J.~C.}\ \bibnamefont {Cooley}}, \bibinfo {author} {\bibfnamefont {W.~L.}\
  \bibnamefont {Hults}}, \bibinfo {author} {\bibfnamefont {L.}~\bibnamefont
  {Morales}}, \bibinfo {author} {\bibfnamefont {D.~J.}\ \bibnamefont {Thoma}},
  \bibinfo {author} {\bibfnamefont {J.~L.}\ \bibnamefont {Smith}}, \bibinfo
  {author} {\bibfnamefont {J.}~\bibnamefont {Boerio-Goates}}, \bibinfo {author}
  {\bibfnamefont {B.~F.}\ \bibnamefont {Woodfield}}, \bibinfo {author}
  {\bibfnamefont {G.~R.}\ \bibnamefont {Stewart}}, \bibinfo {author}
  {\bibfnamefont {R.~A.}\ \bibnamefont {Fisher}},\ and\ \bibinfo {author}
  {\bibfnamefont {N.~E.}\ \bibnamefont {Phillips}},\ }\href
  {http://dx.doi.org/10.1016/S0011-2275(03)00092-4} {\bibfield  {journal}
  {\bibinfo  {journal} {Cryogenics}\ }\textbf {\bibinfo {volume} {43}},\
  \bibinfo {pages} {369} (\bibinfo {year} {2003})}\BibitemShut {NoStop}%
\bibitem [{\citenamefont {Plimpton}(1995)}]{Plimpton1995}%
  \BibitemOpen
  \bibfield  {author} {\bibinfo {author} {\bibfnamefont {S.}~\bibnamefont
  {Plimpton}},\ }\href {\doibase 10.1006/jcph.1995.1039} {\bibfield  {journal}
  {\bibinfo  {journal} {J. Comp. Phys.}\ }\textbf {\bibinfo {volume} {117}},\
  \bibinfo {pages} {1} (\bibinfo {year} {1995})},\ \bibinfo {note}
  {\url{http://lammps.sandia.gov/}}\BibitemShut {NoStop}%
\bibitem [{\citenamefont {Mendelev}\ \emph {et~al.}(2009)\citenamefont
  {Mendelev}, \citenamefont {Kramer}, \citenamefont {Ott}, \citenamefont
  {Sordelet}, \citenamefont {Yagodin},\ and\ \citenamefont
  {Popel}}]{Mendelev2009}%
  \BibitemOpen
  \bibfield  {author} {\bibinfo {author} {\bibfnamefont {M.~I.}\ \bibnamefont
  {Mendelev}}, \bibinfo {author} {\bibfnamefont {M.~J.}\ \bibnamefont
  {Kramer}}, \bibinfo {author} {\bibfnamefont {R.~T.}\ \bibnamefont {Ott}},
  \bibinfo {author} {\bibfnamefont {D.~J.}\ \bibnamefont {Sordelet}}, \bibinfo
  {author} {\bibfnamefont {D.}~\bibnamefont {Yagodin}},\ and\ \bibinfo
  {author} {\bibfnamefont {P.}~\bibnamefont {Popel}},\ }\href {\doibase
  10.1080/14786430902832773} {\bibfield  {journal} {\bibinfo  {journal}
  {Philos. Mag.}\ }\textbf {\bibinfo {volume} {89}},\ \bibinfo {pages} {967}
  (\bibinfo {year} {2009})}\BibitemShut {NoStop}%
\bibitem [{\citenamefont {Shimizu}\ \emph {et~al.}(2007)\citenamefont
  {Shimizu}, \citenamefont {Ogata},\ and\ \citenamefont {Li}}]{Shimizu2007}%
  \BibitemOpen
  \bibfield  {author} {\bibinfo {author} {\bibfnamefont {F.}~\bibnamefont
  {Shimizu}}, \bibinfo {author} {\bibfnamefont {S.}~\bibnamefont {Ogata}},\
  and\ \bibinfo {author} {\bibfnamefont {J.}~\bibnamefont {Li}},\ }\href
  {http://dx.doi.org/10.2320/matertrans.MJ200769} {\bibfield  {journal}
  {\bibinfo  {journal} {Mater. Trans.}\ }\textbf {\bibinfo {volume} {48}},\
  \bibinfo {pages} {2923} (\bibinfo {year} {2007})}\BibitemShut {NoStop}%
\bibitem [{\citenamefont {Stukowski}(2010)}]{Stukowski2010}%
  \BibitemOpen
  \bibfield  {author} {\bibinfo {author} {\bibfnamefont {A.}~\bibnamefont
  {Stukowski}},\ }\href {http://dx.doi.org/10.1088/0965-0393/18/1/015012}
  {\bibfield  {journal} {\bibinfo  {journal} {Model. Simul. Mater. Sci. Eng.}\
  }\textbf {\bibinfo {volume} {18}},\ \bibinfo {pages} {015012} (\bibinfo
  {year} {2010})},\ \bibinfo {note} {\url{http://ovito.org/}}\BibitemShut
  {NoStop}%
\bibitem [{\citenamefont {Dickey}\ and\ \citenamefont
  {Paskin}(1969)}]{Dickey1969}%
  \BibitemOpen
  \bibfield  {author} {\bibinfo {author} {\bibfnamefont {J.~M.}\ \bibnamefont
  {Dickey}}\ and\ \bibinfo {author} {\bibfnamefont {A.}~\bibnamefont
  {Paskin}},\ }\href {\doibase 10.1103/PhysRev.188.1407} {\bibfield  {journal}
  {\bibinfo  {journal} {Phys. Rev.}\ }\textbf {\bibinfo {volume} {188}},\
  \bibinfo {pages} {1407} (\bibinfo {year} {1969})}\BibitemShut {NoStop}%
\bibitem [{\citenamefont {Pathria}(1996)}]{Pathria1996}%
  \BibitemOpen
  \bibfield  {author} {\bibinfo {author} {\bibfnamefont {R.~K.}\ \bibnamefont
  {Pathria}},\ }\href@noop {} {\emph {\bibinfo {title} {Statistical
  Mechanics}}},\ \bibinfo {edition} {2nd}\ ed.\ (\bibinfo  {publisher}
  {Elsevier Butterworth-Heinemann},\ \bibinfo {year} {1996})\BibitemShut
  {NoStop}%
\bibitem [{\citenamefont {R\"osner}\ \emph {et~al.}(2013)\citenamefont
  {R\"osner}, \citenamefont {K\"ubel}, \citenamefont {Peterlechner},\ and\
  \citenamefont {Wilde}}]{RoesnerShearBands}%
  \BibitemOpen
  \bibfield  {author} {\bibinfo {author} {\bibfnamefont {H.}~\bibnamefont
  {R\"osner}}, \bibinfo {author} {\bibfnamefont {C.}~\bibnamefont {K\"ubel}},
  \bibinfo {author} {\bibfnamefont {M.}~\bibnamefont {Peterlechner}},\ and\
  \bibinfo {author} {\bibfnamefont {G.}~\bibnamefont {Wilde}},\ }\href@noop {}
  {\bibfield  {journal} {\bibinfo  {journal} {Ultramicroscopy}\ }\textbf
  {\bibinfo {volume} {submitted}} (\bibinfo {year} {2013})}\BibitemShut
  {NoStop}%
\bibitem [{\citenamefont {Bokeloh}\ \emph {et~al.}(2011)\citenamefont
  {Bokeloh}, \citenamefont {Divinski}, \citenamefont {Reglitz},\ and\
  \citenamefont {Wilde}}]{PhysRevLett.107.235503}%
  \BibitemOpen
  \bibfield  {author} {\bibinfo {author} {\bibfnamefont {J.}~\bibnamefont
  {Bokeloh}}, \bibinfo {author} {\bibfnamefont {S.~V.}\ \bibnamefont
  {Divinski}}, \bibinfo {author} {\bibfnamefont {G.}~\bibnamefont {Reglitz}},\
  and\ \bibinfo {author} {\bibfnamefont {G.}~\bibnamefont {Wilde}},\ }\href
  {http://dx.doi.org/10.1103/PhysRevLett.107.235503} {\bibfield  {journal}
  {\bibinfo  {journal} {Phys. Rev. Lett.}\ }\textbf {\bibinfo {volume} {107}},\
  \bibinfo {pages} {235503} (\bibinfo {year} {2011})}\BibitemShut {NoStop}%
\bibitem [{\citenamefont {Jiang}\ \emph {et~al.}(2005)\citenamefont {Jiang},
  \citenamefont {Pinkerton},\ and\ \citenamefont {Atzmon}}]{Jiang20053469}%
  \BibitemOpen
  \bibfield  {author} {\bibinfo {author} {\bibfnamefont {W.~H.}\ \bibnamefont
  {Jiang}}, \bibinfo {author} {\bibfnamefont {F.~E.}\ \bibnamefont
  {Pinkerton}},\ and\ \bibinfo {author} {\bibfnamefont {M.}~\bibnamefont
  {Atzmon}},\ }\href {http://dx.doi.org/10.1016/j.actamat.2005.04.003}
  {\bibfield  {journal} {\bibinfo  {journal} {Acta Mater.}\ }\textbf {\bibinfo
  {volume} {53}},\ \bibinfo {pages} {3469} (\bibinfo {year}
  {2005})}\BibitemShut {NoStop}%
\bibitem [{\citenamefont {Xie}\ and\ \citenamefont
  {George}(2008)}]{Xie20085202}%
  \BibitemOpen
  \bibfield  {author} {\bibinfo {author} {\bibfnamefont {S.}~\bibnamefont
  {Xie}}\ and\ \bibinfo {author} {\bibfnamefont {E.~P.}\ \bibnamefont
  {George}},\ }\href {http://dx.doi.org/10.1016/j.actamat.2008.07.009}
  {\bibfield  {journal} {\bibinfo  {journal} {Acta Mater.}\ }\textbf {\bibinfo
  {volume} {56}},\ \bibinfo {pages} {5202} (\bibinfo {year}
  {2008})}\BibitemShut {NoStop}%
\bibitem [{\citenamefont {Ritter}\ and\ \citenamefont
  {Albe}(2011)}]{Ritter20117082}%
  \BibitemOpen
  \bibfield  {author} {\bibinfo {author} {\bibfnamefont {Y.}~\bibnamefont
  {Ritter}}\ and\ \bibinfo {author} {\bibfnamefont {K.}~\bibnamefont {Albe}},\
  }\href {http://dx.doi.org/10.1016/j.actamat.2011.07.063} {\bibfield
  {journal} {\bibinfo  {journal} {Acta Mater.}\ }\textbf {\bibinfo {volume}
  {59}},\ \bibinfo {pages} {7082} (\bibinfo {year} {2011})}\BibitemShut
  {NoStop}%
\bibitem [{\citenamefont {Kanungo}\ \emph {et~al.}(2004)\citenamefont
  {Kanungo}, \citenamefont {Glade}, \citenamefont {Asoka-Kumar},\ and\
  \citenamefont {Flores}}]{Kanungo20041073}%
  \BibitemOpen
  \bibfield  {author} {\bibinfo {author} {\bibfnamefont {B.~P.}\ \bibnamefont
  {Kanungo}}, \bibinfo {author} {\bibfnamefont {S.~C.}\ \bibnamefont {Glade}},
  \bibinfo {author} {\bibfnamefont {P.}~\bibnamefont {Asoka-Kumar}},\ and\
  \bibinfo {author} {\bibfnamefont {K.~M.}\ \bibnamefont {Flores}},\ }\href
  {http://dx.doi.org/10.1016/j.intermet.2004.04.033} {\bibfield  {journal}
  {\bibinfo  {journal} {Intermetallics}\ }\textbf {\bibinfo {volume} {12}},\
  \bibinfo {pages} {1073} (\bibinfo {year} {2004})}\BibitemShut {NoStop}%
\bibitem [{\citenamefont {Heuer}(2008)}]{0953-8984-20-37-373101}%
  \BibitemOpen
  \bibfield  {author} {\bibinfo {author} {\bibfnamefont {A.}~\bibnamefont
  {Heuer}},\ }\href {http://dx.doi.org/10.1088/0953-8984/20/37/373101}
  {\bibfield  {journal} {\bibinfo  {journal} {J. Phys.: Condens. Mat.}\
  }\textbf {\bibinfo {volume} {20}},\ \bibinfo {pages} {373101} (\bibinfo
  {year} {2008})}\BibitemShut {NoStop}%
\end{thebibliography}


\end{document}